\documentstyle[epsfig]{elsart}
\begin{document}
\begin{frontmatter}
\title{A two-step optimized measurement for the phase-shift}
\author[dubna,adlershof]{Alexei V. Chizhov}~
\author[modena,dahlem]{Valentina De Renzi}~ 
\author[pavia,adlershof]{Matteo G. A. Paris}
\address[dubna]{Bogolubov~Laboratory~of~Theoretical~Physics \\ 
Joint~Institute~for~Nuclear~Research --- 141980 Dubna, Moscow Reg., Russia}
\address[adlershof]{Arbeitsgruppe 'Nichtklassiche Strahlung' der
Max-Planck-Gesellschaft \\ Rudower Chausse 5, 12489 Berlin, Germany}
\address[modena]{Dipartimento di Fisica dell' Universit\'a di Modena \\ 
and I.N.F.M. -- Unit\'a di Modena, via Campi 113/A, 41100 Modena, Italy}
\address[dahlem]{Fritz-Haber Institut der Max-Planck-Gesellschaft Abteilung 
Physikalische Chemie \\ Faradayweg 4/6, 14195 Berlin, Germany}
\address[pavia]{Dipartimento di Fisica ``Alessandro Volta" dell'Universit\'a
degli Studi di Pavia \\ and I.N.F.M -- Unit\'a di Pavia, via A. Bassi 6, 
I-27100 Pavia, Italy}         
\begin{abstract}
A two-step detection strategy is suggested for the precise measurement
of the optical phase-shift. In the first step an {\em unsharp}, however,
{\em unbiased} joint measurement of the phase and photon number is performed
by heterodyning the signal field. Information coming from this step is then
used for suitable squeezing of the probe mode to obtain a sharp
phase distribution. Application to squeezed states leads to a phase
sensitivity scaling as $\Delta\varphi\simeq N^{-1}$ relative to the
total number of photons impinged into the apparatus. Numerical 
simulations of the whole detection strategy are also also presented.
\end{abstract}
\end{frontmatter}
\section{Introduction}
As a matter of fact, no hermitian operator describing the optical
phase can be defined on the sole Hilbert space of a single mode
radiation field. Nonetheless, measurements of the phase-shift have
been experimentally carried out for quantized field \cite{buc,nfm}.
Moreover, different experimental setups produce different phase
distributions when investigating the same state of radiation \cite{man}.
The contradiction among these facts is only apparent.
Indeed, what is actually measured in physical experiments is always a
phase difference between the signal mode and a reference mode, which
represents the probe of the measuring device.
This probe mode is also a quantized field, characterized by its own
field (phase and amplitude) fluctuations. Therefore, it appears rather
obvious that the resulting phase distribution could show dramatically
different features upon different probe modes. In this paper we are going 
to take advantage of this
fact in order to obtain an optimized measurement for the phase-shift. 
This means a detection scheme leading to a phase distribution as sharp as 
possible, provided that the physical constraint of a fixed amount of energy 
impinged into the apparatus is satisfied. \par
In the next section we briefly review the main features of generalized
phase-space functions, whereas the two-step measurement scheme is
analyzed in details in Section \ref{s:two}. In Section \ref{s:sim}
a numerical simulation of the whole detection strategy is presented, 
in order to confirm the effectiveness of the method also for low excited
states. Section \ref{s:out} closes the paper with some concluding
remarks.
\section{Measuring generalized phase-space distributions}
We are considering here a generic two-photocurrent device, namely an 
apparatus jointly measuring the real $\widehat Z_1$ and the imaginary 
$\widehat Z_2$ parts of the complex photocurrent $\widehat Z= \hat a + 
\hat b^{\dag}$. The operators $\hat a$ and $\hat b$ describe two single 
modes of the radiation  field. We refer to $a$ as the signal mode and to 
$b$ as the probe mode of the device. 
Such kind of devices are readily available in quantum optics. Examples are 
provided by the heterodyne detectors \cite{het}, the eight-port homodyne 
detectors \cite{8po} and the recently introduced six-port homodyne detectors 
\cite{tri}. 
\par 
Each random experimental outcome is represented by a couple of real numbers 
$(z_1,z_2)$ which can be viewed as a complex number $z$ on the plane of the 
field amplitude (phase-space) \cite{rip} .
These are distributed according to a generalized phase space
distribution \cite{wod,buz}
\begin{eqnarray} 
K_b(\alpha,\bar\alpha) = \int_{\bf C} 
\frac{d^2\gamma}{\pi^2}\:  e^{\bar{\gamma}\alpha-\gamma\bar\alpha}\:\Xi 
(\gamma,\bar\gamma) \label{FT}\:, 
\end{eqnarray} 
which is the  Fourier transform of the characteristic function 
\begin{eqnarray} 
\Xi(\gamma,\bar\gamma)=\hbox{Tr}\left\{\hat\varrho\:\exp\left[\bar\gamma
\widehat{Z}-\gamma\widehat{Z}^{\dag}\right]\right\}\label{Xi}\:,
\end{eqnarray}
$\hat\varrho$ being the global density matrix describing both the modes $a$
and $b$. Here, we consider the probe mode to be independent on the signal 
mode, so that the input mode is factorized as $\hat\varrho=\hat\varrho_a
\otimes\hat\varrho_b$. In this case the characteristic function
$\Xi(\gamma,\bar\gamma)$ can be written as a product
\begin{eqnarray}
\Xi (\gamma,\bar\gamma)= \hbox{Tr}\left\{\hat\varrho_a \otimes
\hat\varrho_b \;\; \hat D_a(\gamma) \otimes \hat D_b (-\gamma) \right\}
= \chi_a(\gamma,\bar{\gamma})\;\chi_b(-\gamma,-\bar{\gamma})
\label{pXi}\:,
\end{eqnarray}
being $\hat D_a(\gamma) = \exp\left[ \gamma \hat a^{\dag} -
\bar{\gamma}\hat a \right]$ the displacement operator and 
\begin{equation}
\chi_i(\gamma,
\bar\gamma)= \hbox{Tr}\left\{\hat\varrho\:\hat D_i(\gamma)\right\}
\qquad i=a,b \nonumber\, ;
\end{equation}
the single-mode characteristics function. The latter enters in the definition
of the Wigner function of a single-mode radiation field
\begin{eqnarray}
W_i(\alpha,\bar\alpha )=\int_{\bf C}\frac{d^2\lambda}{\pi}\:\chi_i(\lambda,
\bar{\lambda})\:\exp\left\{\bar\lambda \alpha -\lambda \bar\alpha\right\}
\qquad i=a,b
\label{Wdf}\;.
\end{eqnarray}
We now insert Eq.~(\ref{pXi}) into Eq.~(\ref{FT}). By means of Eq.~(\ref{Wdf})
and using the convolution theorem, we arrive at the result
\begin{eqnarray}
K_b(\alpha,\bar\alpha ) &=& W_a(\alpha,\bar\alpha )\star W_b(-\alpha,-
\bar\alpha ) = \nonumber \\
&=& \int_{\bf C} \frac{d^2 \beta}{\pi^2} \; W_a(\alpha+\beta,
\bar\alpha +\bar\beta )\: W_b(\beta,\bar\beta ) 
\label{Con}
\end{eqnarray}
the symbol $\star$ denoting convolution.
>From Eq.~(\ref{Con}) it results that two-photo\-current devices
allows filtering of the signal Wigner function according to the
probe Wigner function. Therefore, they are powerful apparatus in
order to manipulate and redirect quantum fluctuations.
\section{A two-step measurement scheme for the phase-shift}\label{s:two}
\begin{figure}[ht]
\psfig{file=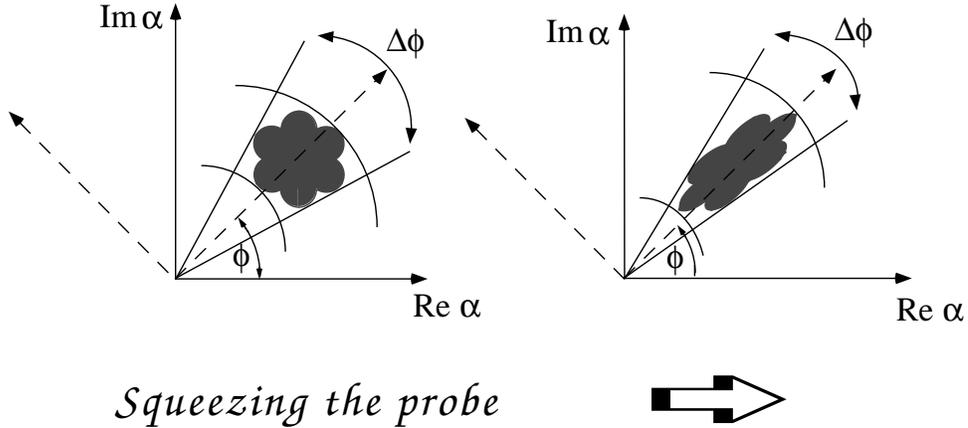,width=14cm}
\caption{\sc Manipulation of quantum fluctuations by squeezing of the
probe mode. In (a) we show a generic, irregularly shaped, $Q$-function
obtainable by two-photocurrent devices with vacuum probe mode. In (b) 
we show the distribution as obtained after squeezing the probe mode in
the direction individuated by the mean value $\phi$ of the signal phase.}
\label{f:show}
\end{figure}

The phase distribution in a two-photocurrent measurement scheme is
defined as the marginal distribution of $K_b(\alpha ,\bar{\alpha})$ 
integrated over the radius,
\begin{eqnarray}
p(\varphi )= \int_0^{\infty} \! \rho\: d\rho\; K_b(\rho e^{i\varphi}, \rho
e^{-i\varphi}) \label{pfib}\:.
\end{eqnarray}
When the probe mode is left unexcited $\hat\varrho_b=|0\rangle\langle
0|$, the probability distribution $K_b(\alpha ,\bar{\alpha})$ coincides
with the customary Husimi $Q$-function
$Q(\alpha,\bar\alpha)=1/\pi \langle\alpha |\hat\varrho |\alpha\rangle$
of the signal mode.
The resulting marginal phase distribution, as defined by Eq.~(\ref{pfib}),
is given by \cite{paul,tangan}
\begin{eqnarray}
p_Q(\varphi ) &=& \int_0^{\infty} \! \rho\:d\rho\; Q(\rho e^{i\varphi},
\rho e^{-i\varphi}) \nonumber \\
              &=& \frac{1}{2\pi} \sum_{n,m}\:\frac{\Gamma
[1+\frac{m+n}{2}]}{\sqrt{n! m!}}\:\exp\{i(n-m)\varphi\}\:
\langle n|\hat\varrho |m\rangle \label{pfiQ}\:.
\end{eqnarray}
The probability $p_Q(\varphi )$ is an unsharp, however unbiased phase 
distribution. That is, it provides a reliable mean value for the phase but 
it is a broad distribution due to the intrinsic quantum noise introduced by 
the joint measurement \cite{yue}. The basic idea in the present 
{\em two-step} scheme is to use information coming from the measurement 
of $p_Q(\varphi )$ in order to suitably squeeze the probe mode in the 
subsequent measurement.
In this way the noise is redirected to the 'useless' direction of
$K_b(\alpha ,\bar{\alpha})$ resulting in the sharper phase distribution.
This procedure is illustrated in Fig.~\ref{f:show} for a generic quantum 
state.
\par
The natural choice for the state to apply this procedure is that of squeezed
states
\begin{eqnarray}
|\alpha,\zeta\rangle=\hat D(\alpha ) \hat S(\zeta)|0\rangle
\label{squeezed}\:,\end{eqnarray}
being $\hat S(\zeta )=\exp\{1/2 (\bar\zeta\hat a^2-\zeta\hat a^{\dag 2})\}$
the squeezing operator with complex parameter $\zeta= r \exp\{2i\psi\}$. 
Squeezed states, in fact, show phase-dependent
field fluctuations and can be presently produced with reliable
experimental techniques. The mean photon number is given by
$\langle\hat n\rangle\equiv N=|\alpha|^2 + \sinh^2 r \equiv N_{coh}
+ N_{sq}$, where the coherent and squeezing contributions can be
clearly distinguished. Squeezed states have been largely considered in
interferometry, usually leading to high-precision
measurements, though only for a special value of the phase-shift 
(the so-called working point of the interferometer) \cite{cav,par}.
Without loss of generality, we consider here a squeezed
state with both coherent and squeezing phases equal to zero.
This is accomplished by choosing $\alpha=x_s \in {\bf R}$ and
$\zeta=r_s \in {\bf R}$.
\par
In the first step of the measurement we leave the probe unexcited. The
experimental outcomes are thus distributed according to the Husimi
$Q$-function of a squeezed state which is a double Gaussian given by
\begin{eqnarray}
Q(\alpha,\bar\alpha) = \frac{1}{2\pi\sigma_1\sigma_2}
\exp\left\{-\frac{[\hbox{Re}(\alpha)-x_s]^2}{2\sigma_1^2}
           -\frac{[\hbox{Im}(\alpha)]^2}{2\sigma_2^2}\right\},\\
\sigma_1^2=\frac{1}{4}\left(1+\exp\{2r_s\}\right), \qquad\qquad
\sigma_2^2=\frac{1}{4}\left(1+\exp\{-2r_s\}\right)
\label{Q}\:.
\end{eqnarray}
The marginal phase distribution $p_Q(\varphi)$ reads as follow
\begin{eqnarray}
p_Q(\varphi)=\frac{1}{2\pi\mu\cosh r_s}\:e^{-\frac{x_s^2}{2\sigma_1^2}}
\left\{
1+\sqrt{\pi}\:\frac{\nu}{\sqrt{\mu}}\:e^{\frac{\nu^2}{\mu}}
\left[1+\hbox{Erf}\left(\frac{\nu}{\sqrt{\mu}}\right)\right]\right\}
\label{Qphi}\:,
\end{eqnarray}
where $\hbox{Erf}(x)=2/ \sqrt{\pi}\int_0^x dt \exp\{-t^2\}$ denotes
the error function and
\begin{eqnarray}
\mu=\frac{1}{2}\left(\frac{\cos^2\varphi}{\sigma_1^2}
+\frac{\sin^2\varphi}{\sigma_2^2}\right),\qquad\qquad
\nu=\frac{x_s\cos\varphi}{2\sigma_1^2}
\label{par}\:.
\end{eqnarray}
For the large signal intensity ($x_s \gg 1$), it is possible to expand
$p_Q(\varphi)$ up to the second order in $\varphi$. The resulting
distribution is a Gaussian
\begin{eqnarray}
p_Q(\varphi)=\frac{1}{\sqrt{2\pi}\Delta_{\varphi}}
\exp\left\{-\frac{\varphi^2}{2\Delta_{\varphi}^2}\right\},
\qquad\qquad
\Delta_{\varphi}=\frac{\sigma_2}{x_s}
\label{fgau}\:.
\end{eqnarray}
In the case of the highly squeezed signal mode ($r_s \gg 1$), the width of 
the phase distribution~(\ref{fgau}) turns out to be
\begin{equation}
\Delta_{\varphi}=\frac{1}{2\sqrt{\beta_s N}}
\label{wid}\:,
\end{equation}
being $\beta_s = x_s^2/N \equiv N_{coh}/N$ the coherent fraction of the total
number of photons.  The rms variance in Eq.~(\ref{fgau}) is a measure of the
precision in the phase measurement, namely, the sensitivity in revealing
phase fluctuations. Eq.~(\ref{wid}) indicates that the phase distribution
with an unexcited probe is broadened (unsharp) as the scaling $\Delta\varphi
\propto N^{-1/2}$ is distinctive of coherent (semiclassical) interferometry. 
Nonetheless,
reliable information on the mean-phase value can still be extracted from
$p_Q(\varphi)$.  Indeed, the second step of the measurement is performed with
the probe mode excited to a squeezed vacuum $|\zeta\rangle$ whose phase is
matched to that extracted from the first measurement step.  Therefore, the 
outcome
probability distribution becomes a squeezed $Q$-function.  For the squeezed
state of Eq.~(\ref{squeezed}) this is still a double Gaussian on the complex
plane.  However, the variances are now given by
\begin{eqnarray}
\sigma_1^2 &=& \frac{1}{4}\left[\left(\cosh 2r_p - \sinh 2r_p \cos 2\psi_p
\right)+\exp\{2r_s\}\right], \nonumber \\
\sigma_2^2 &=& \frac{1}{4}\left[\left(\cosh 2r_p - \sinh 2r_p \cos 2\psi_p
\right)+\exp\{-2r_s\}\right]
\label{K}\:,
\end{eqnarray}
where $r_p$ is the squeezing parameter of the probe mode and $\psi_p$ stands
for its phase. The latter is chosen equal to the mean signal phase
$\bar\varphi$ extracted from the first step of the measurement.
The marginal phase probability $p_{\zeta} (\varphi)$ has the same
complicated structure of $p_Q(\varphi)$ in Eq.~(\ref{Qphi}). For the large
signal intensity ($x_s \gg 1$) and high squeezing of the signal and probe 
($r_s\gg r_p\gg 1$), it is well approximated by a Gaussian with rms variance 
given by 
\begin{eqnarray} \Delta\varphi=\frac{1}{4\sqrt{\beta_s\beta_p} N} 
\label{opt}\:,
\end{eqnarray}
being $N$ the total mean photon number impinged into the apparatus (signal
plus probe). The improvement in the precision is apparent.  In
Eq.~(\ref{opt}) $\beta_s$ and $\beta_p$ denote the coherent and squeezing
energy fraction, respectively, of the signal and probe, $\beta_s= x_s^2/N$,
$\beta_p=\sinh^2 r_p/N$, relative to the total number of photons (signal 
plus probe) impinged into the apparatus. \par
\section{Low excited states: numerical simulations}\label{s:sim}
\begin{figure}[ht]
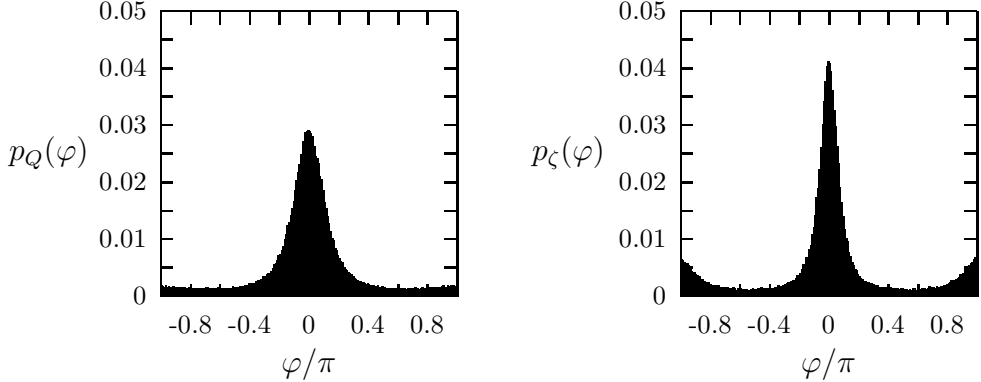

\begin{tabular}{cc}
\input twostep_f2a.tex & \input twostep_f2b.tex
\end{tabular}
\caption{\sc Two-step measurement of the phase of a squeezed
state. In (a): the phase histogram from a simulated experiment with
an unexcited probe. In (b): the phase histogram from a simulated experiment
with a probe excited to a squeezed vacuum whose phase is matched to the mean
phase extracted from (a). The total mean photon number impinged into the
apparatus is $N=2$ in both cases. The simulated experimental sample was
composed by $10^5$ data, whereas $200$ bins were chosen for the phase
histograms.
\label{f:sim}}
\end{figure}

In order to show the effectiveness of the present procedure also for low
excited states, we have performed numerical simulations of the whole detection
scheme.  In Fig.~\ref{f:sim} the two-step phase distributions are shown as
coming from a simulated experiment on low excited squeezed states. Each
experimental event in the joint measurement consists of two
photocurrents which in turn can be viewed as a point on the complex plane of
the field amplitude.  The phase value inferred from each event is the polar
angle of the point itself.  The experimental histogram of the phase
distribution is thus obtained by dividing the plane into angular bins and
then counting the number of points which fall into each bin.
In Fig.~\ref{f:sim}a we report the
phase histogram from a simulated two-photocurrent measurement with
an unexcited probe (the first step) of a squeezed state with the given mean
photon number $N=2$ and the squeezing fraction $\sinh^2 r_s/N= 1/4$. The mean
value for the phase obtained from this step appeared to be
$\bar\varphi\simeq 3.3*10^{-3} rad$.  In the second step, the probe is
excited to a squeezed vacuum with squeezing phase phase $\bar\varphi$ and 
squeezing
fraction $\sinh^2 r_p/N=1/4$ relative to the total photon number $N=2$.  The
resulting phase histogram is shown in Fig.~\ref{f:sim}b.  It is
apparently sharper than the first one, even though the signal energy has
been decreased to maintain the same total energy $N=2$ impinged into the 
apparatus. Some tails, due to squeezing, appears around $\varphi=\pm\pi$ in
the second-step distribution. However, this is not dangerous for the
precision of the measurement as they can only be $\pi$-symmetrically placed
relative to the central peak. On the contrary, they can even be used for
further improvement of the measurement sensitivity \cite{sqv}.
\section{Summary and Remarks}\label{s:out}
In conclusion, a two-step optimized phase detection scheme has been
suggested. It uses the possibility to manipulate quantum fluctuations
offered by two-photocurrent devices in order to improve precision. 
In the first step a number $n$ (not too small) of measurements are performed 
in order to accurately determine the mean value for the phase. This value is 
then used to perform the subsequent $n$ measurement in the second step. 
The resulting scheme is much more accurate that simply making $2*n$ 
measurements using the first step setup. One should also notice that the 
error of a measurement in the second step comes from the average over the 
possible values of error resulting from using a particular squeezing phase
obtained in the first step. Therefore, in order to make the present analysis
correct, the number of measurements in both the two steps should be not 
too small. The extreme case, in which just one measurement is made for each 
step, has been analyzed by Wiseman et al \cite{wis} and by D'Ariano et al 
\cite{fed}. In Wiseman' strategies the information obtained during a 
measurement is used to alter the setup continuously, 
whereas in Ref. \cite{fed} the information from a 
single measurement is immediately used to modify the setup for the subsequent 
measurement. On the other hand, 
in the present scheme, the feedback act on blocks of data.
\par
Application to highly excited squeezed states 
lead to high-sensitivity measurement, with phase sensitivity scaling as 
$\Delta\varphi \propto N^{-1}$ 
relative to the total number of photons impinged into the apparatus. 
This result is valid for any value of the phase-shift itself and
represents a crucial improvement with respect to conventional
interferometers, where fluctuations of the phase-shift can be detected
only around some fixed working point. 
The effectiveness of the present two-step procedure has been confirmed 
also for  low excited states by means of numerical simulations of the whole 
detection strategy.


\begin{thebibliography}{99} 
\bibitem{buc} H. Gerhardt, U. Buchler, G Liftin,
Phys. Lett. A{\bf 49}, 119 (1974).
\bibitem{nfm} J.W.Noh, A.Foug\'eres, L.Mandel, Phys. Rev. Lett. {\bf 67}, 1426 
(1991);Phys. Rev. A {\bf 45}, 424 (1992); Phys. Rev. A {\bf 46}, 2840 (1992).
\bibitem{man} Torgeson J R and Mandel L 1996 {\em Phys. Rev. Lett.} {\bf 76} 
3939
\bibitem{het} J. H. Shapiro, S. S. Wagner, IEEE J. Quantum Electron.
QE{\bf20}, 803 (1984); H. P. Yuen, J. H. Shapiro, IEEE Trans. Inform. Theory
IT{\bf 26}, 78 (1980).
\bibitem{8po} N.G. Walker, J.E. Carrol, Opt. Quantum Electr. {\bf 18}, 355
(1986); N. G. Walker, J. Mod. Opt. {\bf 34}, 15 (1987); Y. Lay, H. A. Haus,
Quantum Opt. {\bf 1}, 99 (1989).
\bibitem{tri} M. G. A. Paris, A. Chizhov, O. Steuernagel, Opt. Comm. 
{\bf 134}, 117 (1997).
\bibitem{rip} G. M. D'Ariano, M. G. A. Paris, Phys. Rev. A {\bf 49}, 3022 (1994).
\bibitem{wod} K. Wodkiewicz, Phys. Rev. Lett. {\bf 52}, 1064 (1984); Phys. Lett. A{\bf 115}, 304 (1986); Phys. Lett. A{\bf 129}, 1 (1988). 
\bibitem{buz} V. Buzek, C. H. Keitel, P. L. Knight, Phys. Rev. A{\bf 51}, 2575; Phys. Rev. A{\bf 51}, 2594. 
\bibitem{paul} H.Paul, Fortschr. Phys. {\bf 22}, 657 (1974).
\bibitem{tangan} R.Tana\'s, Ts.Gantsog, Phys. Rev. A {\bf 45}, 5031 (1992).
\bibitem{yue} H. P. Yuen, Phys. Lett. A {\bf 91}, 101 (1982).
\bibitem{cav} C. M. Caves, Phys. Rev. D {\bf 23}, 1693 (1981).
\bibitem{par} M. G. A. Paris, Phys. Lett. A {\bf 201}, 132 (1995).  
\bibitem{sqv} M. G. A. Paris, Mod. Phys.  Lett. B, {\bf 9}, 1141 (1995). 
\bibitem{wis} H. M. Wiseman, Phys. Rev. Lett. {\bf 75} 4587 (1995); 
H. M. Wiseman, R. B. Killip, Phys. Rev. A {\bf 56} 944 (1997). 
\bibitem{fed} G. M. D'Ariano, M. G. A. Paris e R. Seno, Phys. Rev. A  {\bf 54}, 
4495 (1996).
\end{thebibliography}
\end{document}